\documentclass[aps,superscriptaddress,prl]{revtex4}
\usepackage{graphicx,amsfonts,amsmath,color,amsbsy,amssymb}

\begin{document}
\title{Mathematical Modelling of Heart Rate Changes in the Mouse*}

\author{Mark Christie}
\affiliation{Institute of Pharmaceutical Science, King's College London, London SE1 9NH, U.K.}

\author{Manasi Nandi}
\affiliation{Institute of Pharmaceutical Science, King's College London, London SE1 9NH, U.K.}

\author{Yanika Borg}
\affiliation{Department of Mathematics and Department of Biochemical Engineering, University College London, London WC1E 6BT, U.K.}

\author{Valentina Carapella}
\affiliation{Department of Computer Science, University of Oxford, Oxford OX1 3QD, U.K.}

\author{Gary Mirams}
\affiliation{Department of Computer Science, University of Oxford, Oxford OX1 3QD, U.K.}

\author{Philip Aston}
\affiliation{Department of Mathematics, University of Surrey, Guildford GU2 7XH, U.K.}

\author{Saziye Bayram}
\affiliation{Department of Mathematics, SUNY Buffalo State, Buffalo NY 14222, U.S.A.}

\author{Radostin Simitev}
\affiliation{Department of Mathematics and Statistics, University of Glasgow, Glasgow G12 8QW, U.K.}

\author{Jennifer Siggers}
\affiliation{Department of Bioengineering, Imperial College, London, U.K.}

\author{Buddhapriya Chakrabarti}
\affiliation{Department of Mathematical Sciences, Durham University, Durham, DH1 3LE, U.K.}

\date{\today}

* This report summarises the outcomes from the UK MMSG NC3R's Study Group meeting, 15$^{\rm th}$--18$^{\rm th}$ April 2013 in response to a problem entitled `Modelling heart rate changes in the mouse as a series of delayed, weakly coupled oscillators', presented by MC and MN.

\begin{abstract}
The CVS is composed of numerous interacting and dynamically regulated physiological subsystems which each generate measurable periodic components such that the CVS can itself be presented as a system of weakly coupled oscillators.  The interactions between these oscillators generate a chaotic blood pressure waveform signal, where periods of apparent rhythmicity are punctuated by asynchronous behaviour. It is this variability which seems to characterise the normal state.
We used a standard experimental data set for the purposes of analysis and modelling. Arterial blood pressure waveform data was collected from conscious mice instrumented with radiotelemetry devices over $24$ hours, at a $100$Hz and $1$kHz time base. During a $24$ hour period, these mice display diurnal variation leading to changes in the cardiovascular waveform.
We undertook preliminary analysis of our data using Fourier transforms and subsequently applied a series of both linear and nonlinear mathematical approaches in parallel. We provide a minimalistic linear and nonlinear coupled oscillator model and employed spectral and Hilbert analysis as well as a phase plane analysis.
This provides a route to a three way synergistic investigation of the original blood pressure data by a combination of physiological experiments, data analysis \textit{viz.}\ Fourier and Hilbert transforms and attractor reconstructions, and numerical solutions of linear and nonlinear coupled oscillator models. We believe that a minimal model of coupled oscillator models that quantitatively describes the complex physiological data could be developed via such a method. Further investigations of each of these techniques will be explored in separate publications.
\end{abstract}


\maketitle

\section{Introduction}
\label{intro}

The cardiovascular system (CVS) has evolved to keep blood in continuous motion such that adequate cellular oxygen and nutrient requirements are met at any given time. The system must be able to adapt to acute changes in the body's physiology such as sleep, postural changes and exercise~\cite{Dampney:02}.

The CVS is composed of numerous interacting and dynamically regulated physiological subsystems which each generate measurable periodic components~\cite{Akselrod:81,Stefanovska:01,Stefanovska:01a} such that the CVS can itself be presented as a system of weakly coupled oscillators~\cite{Stefanovska:01, Stefanovska:01a, Stefanovska:07}.

The interaction between these coupled oscillators with the CVS \textit{in vivo} generates a chaotic signal measured as a sinusoidal blood pressure waveform, where periods of apparent rhythmicity are punctuated by asynchronous behaviour. It is this variability which seems to characterise the `normal' state.  Indeed, it has been known for decades that heart rate variability (HRV; the temporal beat to beat variation in electrocardiogram (ECG) recordings) can be used as a marker of `normal' cardiovascular function. Loss of this variability is associated with pathophysiology such as foetal distress~\cite{Hon:63} and is used as a prognostic marker to predict mortality risk in patients post myocardial infarction~\cite{Kleiger:87}.

Similarly, in pathophysiological states such as septicaemia and congestive heart failure, a reduction in HRV is associated with poorer outcome. Some authors~\cite{Godin:96} have described the reduced HRV in septic shock (which occurs in mice and humans) as a decoupling of oscillators although there is evidence that forcing the system to synchrony (\textit{e.g.}\ through mechanical ventilation setting the respiratory rhythm~\cite{Buykin:08}) may have the same effect. In that case it would be the equivalent of the mechanics of a set of forced coupled oscillators.  An interesting outcome of this weakly-coupled oscillator model is that the `normal' condition for HRV seems to reside somewhere between complete synchrony and chaos, possibly because in a conscious animal each of the oscillators in the system receives an occasional increase in power.

In experiments where HRV is analysed in mice it is clear that there is oscillator coupling between respiratory cycles and HR, there is an autonomic sympathetic and parasympathetic component to the control of HR, and that in mouse models of septic shock, reduced HRV precedes the onset of temperature and BP changes~\cite{Baudrie:07,Tsai:12,Fairchild:09}. In all three respects, clinical studies of HRV in normal human subjects parallel the mouse data, with evidence of synchronisation with respiration (referred to as respiratory sinus arrhythmia~\cite{Schafer:98, Buchner:09}, modulation by the sympathetic and parasympathetic nervous systems~\cite{Pomeranz:85} and a tendency to reduced HRV in response to endotoxaemia~\cite{Godin:96}. The latter response has been proposed as an early clinical marker in sepsis~\cite{Barnaby:02,Chen:07,Ahmad:09}, post-stroke infections~\cite{Gunther:12} and multiple organ dysfunction syndrome (end stage septic shock)~\cite{Pontet:03}.


\section{Biological parameter estimates}
\label{bioparameterest}

In experimental animals the sinusoidal blood pressure waveform is commonly reduced to peaks, troughs and amplitude measures that describe systolic, diastolic and pulse pressures respectively, whilst the reciprocal of the beat to beat interval enables calculation of heart rate. This waveform is influenced by physiological subsystems including the respiratory cycle, parasympathetic and sympathetic autonomic outflows which can, in turn, alter the blood pressure waveforms such that they become considerably more complex. Such complexity is often overlooked during conventional analysis and the nature of coupling between physiological subsystems is rarely described.

Experiments that determine long-term changes in BP and HR in mice are usually conducted by implanting a blood pressure transducer/transmitter that allows data to be collected remotely from a conscious, freely moving animal, minimising the confounding effects of stress, as animals are left undisturbed in their home cages. These systems are capable of collecting large amounts of continuous data at high sampling frequencies and over long time periods (days to weeks), but these data series are often irregular, strongly non-stationary and noisy. Conventionally, the data are summarised by averaging, time-binning and filtering (to remove obvious artefacts) and are often presented as discontinuous blocks where, for example, an average BP or HR estimate over a $10$ minute time window is used to represent the output for that hour of recording.  This approach has the disadvantage that a) much of the data is collected but never analysed and b) more subtle interpretations such as the determination of HRV are often 
ignored.

We proposed a move beyond simple descriptors of the variable nature of HR, such as HRV, towards modelling the complexity of HR changes, in the first instance, by reference to a series of weakly coupled oscillators. This model has been applied to neural oscillations providing the basis for modelling of other biological systems~\cite{Stefanovska:01,Stefanovska:01,Ermentrout:1891}.  In the case of the mouse cardiovascular system, some of the oscillators have been described by spectral analysis (\textit{e.g.}\ Fourier transform), the heart has an intrinsic (uncoupled) frequency of around $8$Hz, but oscillates between$\approx 10$Hz and $\approx 7$Hz under the influence of the autonomic nervous system, which has oscillator frequencies of $\approx 0.1$ and $1 $Hz for sympathetic and parasympathetic modulation, respectively. Respiratory oscillations in the conscious mouse occur in the $2-6$Hz frequency range~\cite{Baudrie:07,Tsai:12}. However, it is recognised that Fourier analysis is best suited to a signal that is 
regular, stationary
and without artefact. Furthermore, in a conscious animal the frequency signals are continuous variables, which leads to a broadening of the frequency power band associated with each physiological system.

We used a standard experimental data set for the purposes of analysis and modelling. Arterial blood pressure waveform data was collected from conscious mice instrumented with radiotelemetry devices~\cite{Nandi:12} over $24$ hours, at a $100$Hz and $1$kHz time base. During a $24$ hour period, these mice display diurnal variation leading to changes in cardiovascular waveform data. Generally speaking, they are active during the night (lights off; 8pm--8am) and resting during the day (lights on; 8am--8pm). These data sets were used for subsequent analysis and modelling.

\section{Initial modelling assumptions}
\label{assumptions}

\begin{figure}
\includegraphics[width=8cm]{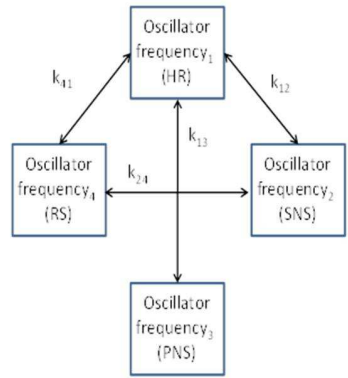}
\caption{The original proposal of treating the cardiovascular system as a set of weakly coupled oscillators, showing various interdependencies. HR = heart rate, SNS = sympathetic nervous system, PNS = parasympathetic nervous system, RS = respiratory system.}\label{Figure1}
\end{figure}

The original proposal involved modelling the cardiovascular system as a series of delayed weakly coupled oscillators is shown in Fig.~\ref{Figure1}.

Whilst the intact biological system will appear to have delays that occur between changes in the subsystems, for simplicity we analysed our model systems without introducing a delay constant, assuming instantaneous propagation of signals across different subsystems $i$ and $j$, parameterised by a single coupling constant $k_{ij}$. Furthermore, in attempting to arrive at a description of a minimal model that captures the full complexity of the system, four physiological oscillators; the heart rate HR; sympathetic nervous system SNS; parasympathetic nervous system PNS; and respiratory rate RR were identified. The likely coupling between each of the oscillators was identified based on the known physiology and interactions between these oscillators.

We undertook preliminary analysis of our data using Fourier transforms in order to compare them with the published literature. We subsequently applied a series of both linear and nonlinear mathematical approaches in parallel. The advantage of this latter approach is that the entire data set is modelled mathematically without \emph{a priori} reference to the physiological state of the animal. Physiological validation of the developed mathematical models was undertaken retrospectively, by inspecting the raw waveform data and inferring which physiological changes were likely to have occurred.

\begin{figure}
\includegraphics[width=8cm]{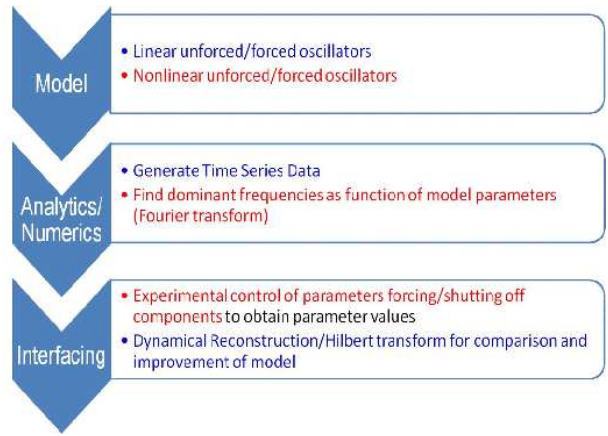}
\caption{The original proposal of treating the cardiovascular system as a set of weakly coupled oscillators.}\label{Figure2}
\end{figure}

A schematic of our workflow is summarised in Fig.~\ref{Figure2}

\section{Mathematical models}
\label{models}

\subsection{Introduction to Mathematical Models}
\label{mathintro}

There exists a substantial body of literature regarding cardiovascular variability which has been more completely reviewed by Karim \textit{et al.}~\cite{Karim:11}, Acharya \textit{et al.}~\cite{RajendraAcharya:06}, Souza Neto \textit{et al.}~\cite{Souza:03}, and Stein \textit{et al.}~\cite{Stein:99}. Recent advances in computational techniques have provided novel tools for fast and accurate signal analysis that has helped advance the field significantly. As a result several mathematical models and techniques that study HRV have been developed over the past three decades. We provide a brief review of these methods in this section.

The issue of whether the HRV is chaotic or not is a highly controversial topic. In 2008, the journal \textit{Chaos} decided to institute a new section in order to address recent and controversial topics in the field of nonlinear science, and dedicated its first section to the topic ``Is the Normal Heart Rate Chaotic?''. The contributed papers to this section ranged from deterministic to stochastic systems (see 
\cite{Alvarez-Ramirez:09,Baillie:09,Buchner:09,Freitas:09,Hu:09,Sassi:09,Wessel:09,Zhang:09}.)

The mathematical methods that have been used in the literature of HRV (mainly for the assessment of cardiovascular signals and their relationship to diseases) may include nonlinear mathematical modelling, spectral analysis (frequency domain), Hilbert transform, wavelet analysis (frequency-time domain), as well as stochastic and statistical modelling.

Spectral analysis, in particular Fourier transforms, is among the most commonly used mathematical techniques. In the 1980's, it was considered to be a modern perspective~\cite{Kay:81}. It is still a valid and powerful technique that provides frequency resolution of stationary signals, but it is known that the cardiovascular associated signals are essentially non-stationary. Hence continuous-time wavelet analysis, which provides time and frequency resolution, of the HRV is now much more commonly employed than Fourier transforms. A compendium of these techniques can be found in~\cite{Addison:05,Pomeranz:85,Akselrod:81,Berger:86,Bracic:98,Busek:05,Guzzetti:88} and ~\cite{RajendraAcharya:05,Singh:04,Faust:04}.

Besides Fourier and wavelet analysis, Hilbert transformations are also used in the HRV literature. A recent paper by Li \textit{et al.}~\cite{Li:11} introduces a modified technique based on the Hilbert-Huang transform (HHT) to improve the spectrum estimates of HRV. The HHT (which is based on the empirical mode decomposition approach to decompose the HRV signal into several monocomponent signals that become analytic signals by means of the Hilbert transform) is proposed to extract the features of preprocessed time series and to characterise the dynamic influence of the parasympathetic and sympathetic nervous systems on the heart.

Some groups, on the other hand, have developed mathematical models involving delayed differential equations to study the HRV phenomenon. The bifurcation theory of nonlinear systems was also used to explain the high sensitivity of the heart rate oscillatory pattern to model parameter changes (see \cite{Cavalcanti:96,Cavalcanti:00}). On the other hand Hu \textit{et al.}~\cite{Hu:09} employed a new multiscale complexity measure, the scale-dependent Lyapunov exponent (SDLE), to characterise HRV. Their analysis showed that the HRV data were mostly stochastic.

Using the language of stochastic processes, Buchner \textit{et al.}~\cite{Buchner:09} analysed the coupled respiratory and heart systems taking the data from polysomnographic recordings of two healthy individuals. Cavalcanti~\cite{Cavalcanti:00} analysed the influence of the arterial baroreflex on HRV by using a mathematical model of heart rate baroreceptor control. The bifurcation theory of nonlinear systems was also used to explain the high sensitivity of the heart rate oscillatory pattern to model parameter changes. Some other mathematical models, deterministic or stochastic, include those described by Kuusela \textit{et al.}~\cite{Kuusela:03}, Buchner \textit{et al.}~\cite{Buchner:09}, Mezentseva \textit{et al.}~\cite{Mezentseva:08} and Karavaev \textit{et al.}~\cite{Karavaev:09}. For statistical modelling one might be interested in the work of Luchinsky \textit{et al.}~\cite{Luchinsky:05}. They present a Bayesian dynamical inference method for characterising cardiorespiratory dynamics in humans by inverse modelling from 
blood pressure time-series data.

Finally, the cardiovascular system has also been studied as a nonlinear dynamical system consisting of coupled oscillators. A fairly recent study by Jamsek \textit{et al.}~\cite{Jamsek:10} proposed a complementary approach combining wavelet bi-spectral analysis~\cite{Jamsek:04} and information theory in order to study the time-varying basic frequencies of the interacting oscillatory systems. This approach allowed them to uncover the interacting properties and reveal the nature, strength, and direction of coupling. Stefanovska \textit{ et al.}~\cite{Stefanovska:01,Stefanovska:01a,Stefanovska:07} introduced a mathematical model of the cardiovascular system where oscillators were coupled with linear couplings. The model was then simulated numerically. Censi~\textit{et al.}~\cite{Censi:02} performed a quantitative study of coupling patterns between respiration and spontaneous rhythms of heart rate and blood pressure variability signals by using Recurrence Quantification Analysis. Unfortunately, some models were
based on a very simplistic description of the cardiovascular system and of the heart. Hence, their value relied on the analysis of the mathematical properties of the equations rather than in physiological reliability.

All of these previous modelling attempts mentioned above underline the need for a comprehensive mathematical model. Motivated by this we provide a minimalistic coupled oscillator (both linear and nonlinear) model; we are not aware of any previous model that summarises all these aspects into a single theoretical structure. We also employ spectral and Hilbert analysis as well as a phase plane analysis. We will explain our models in detail in the next sections.

\subsection{Data Analysis}
\label{analysis}
\begin{figure}
\includegraphics[width=6cm]{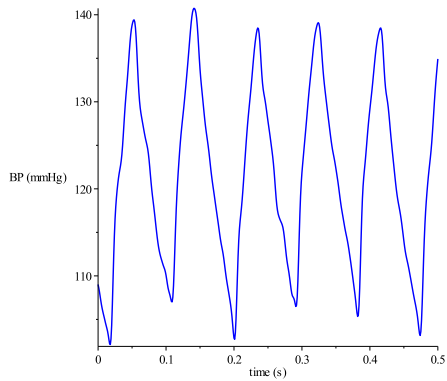}
\caption{A sample of the mouse heart data over $0.5$s.}
\label{data}
\end{figure}

\begin{figure}
\includegraphics[width=8cm]{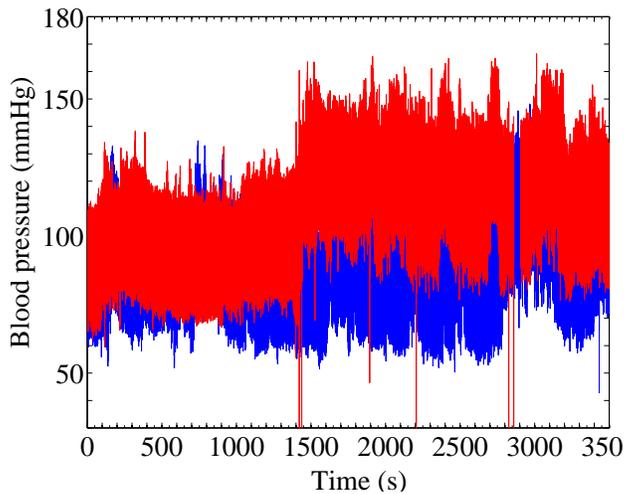}\\
\caption{Representative blood pressure time series data during active (red) and resting (blue) states. Slight difference in the baseline variation can be inferred from the figure.}\label{TimeSeries}
\end{figure}

Blood pressure waveform data was collected over a $24$ hour period~\cite{Nandi:12} from a conscious mouse sampled at $100$Hz and $1$kHz. These cover the resting (8am--8pm) and active (8pm--8am) periods. The challenge in the field has been to develop methods to detect and quantify significant changes in activity based on the blood pressure data. Over a narrow time window ($\approx 0.5$s) the blood pressure data appears to be (approximately) periodic (see Fig.~\ref{data}). However over long time scales the waveform looks aperiodic (see Fig.~\ref{TimeSeries}). This behaviour of short time rhythmicity that is lost over time is the heart rate variability (HRV) discussed above and reviewed in ~\cite{RajendraAcharya:06}.
In an attempt to quantify active and resting states by way of analysing HRV we employ a variety of methods including 
Fourier analysis, a nonlinear analysis,
and Hilbert transforms. We have been able to quantify resting vs.\ active behaviour in a conscious mouse based on these analyses. We outline the results below.

\begin{figure}
\includegraphics[width=7cm]{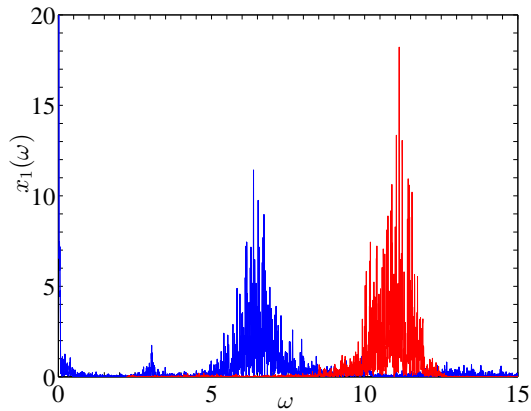}\\
\caption{Fourier transform of the blood pressure data shown in Fig.~\ref{TimeSeries} during the active (red) and resting (blue) phases. Natural frequencies of the individual oscillators have been extracted by fitting a Gaussian waveform to piecewise continuous FFT data.}\label{FFT}
\end{figure}

\subsubsection{Fourier Transform}
\label{FourierT}
\begin{figure}
\includegraphics[width=7cm]{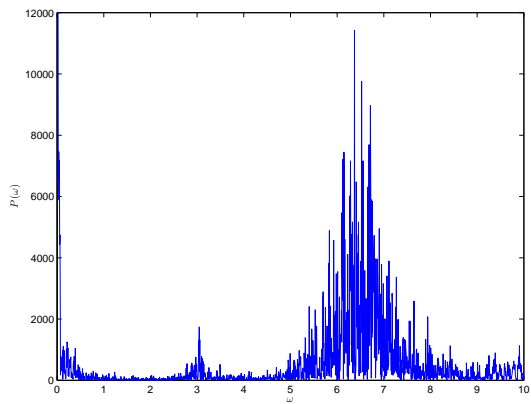}
\caption{FFT applied to 100Hz data set from a mouse during the resting state (1--2pm).}
\label{fig:FFT}
\end{figure}

Fourier analysis was carried out on active 9--10pm and resting 1--2pm data sampled at a frequency of 100Hz. The power spectrum obtained from the Fourier transformed data is shown in Fig.~\ref{fig:FFT}. Three distinct peaks can be observed in the power spectrum data. Taking our cue from the literature values of the frequencies of the different physiological oscillators we considered piecewise data in three different regimes of $\omega$, \textit{i.e.}\ $\delta \omega_{2} \approx 0.02-0.4$Hz, $\delta \omega_{4} \approx 2.85-3.15$Hz and $\delta \omega_{1} \approx 5.7-7.3$Hz.

These data sets were subsequently fitted to a Gaussian function of the form $P(\omega) = \frac{1}{\sqrt{2\pi \sigma}} \exp[ - {\left( \omega - \omega^{*} \right)^{2}} {2 \sigma^2} ]$, using a nonlinear regression scheme to extract the mean value $\omega^{*}$ and the variance $\sigma$ for each section of the data. The mean value corresponds to the natural frequency of each of the physiological subsystems while the variance denotes its spread. Three natural frequencies of the oscillators $\omega_{2} \approx 0.1$Hz (SNS), $\omega_{4} \approx 3.05$Hz (RS) and $\omega_{1} \approx 6.4$Hz (HR) were obtained from this analysis. It is interesting to note that the peak corresponding to the parasympathetic nervous system (PNS) $\omega_{3} \approx 0.7$Hz is not picked up in our Fourier analysis. According to physiological convention, peak frequencies higher than $10$Hz are not usually considered relevant, but it is interesting to note the presence of a strong peak around a value $\approx 2 \omega_{1} \approx 12.5-13$Hz 
range. This may be due to the presence of higher harmonics in the data and will be explored in the future.
\begin{figure}[t]
\centering
\mbox{
{\includegraphics[width=2.2in]{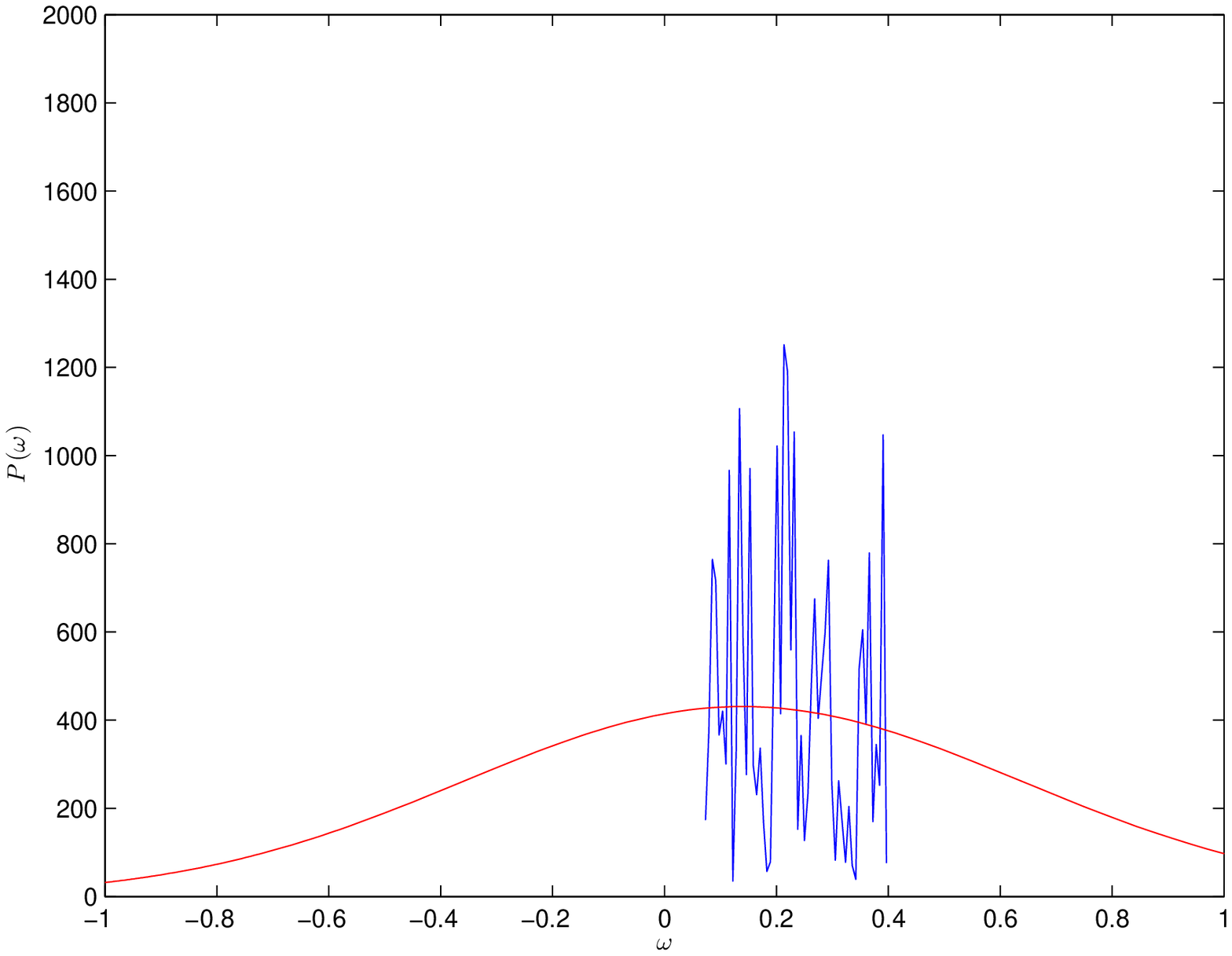}}
\quad
{\includegraphics[width=2.2in]{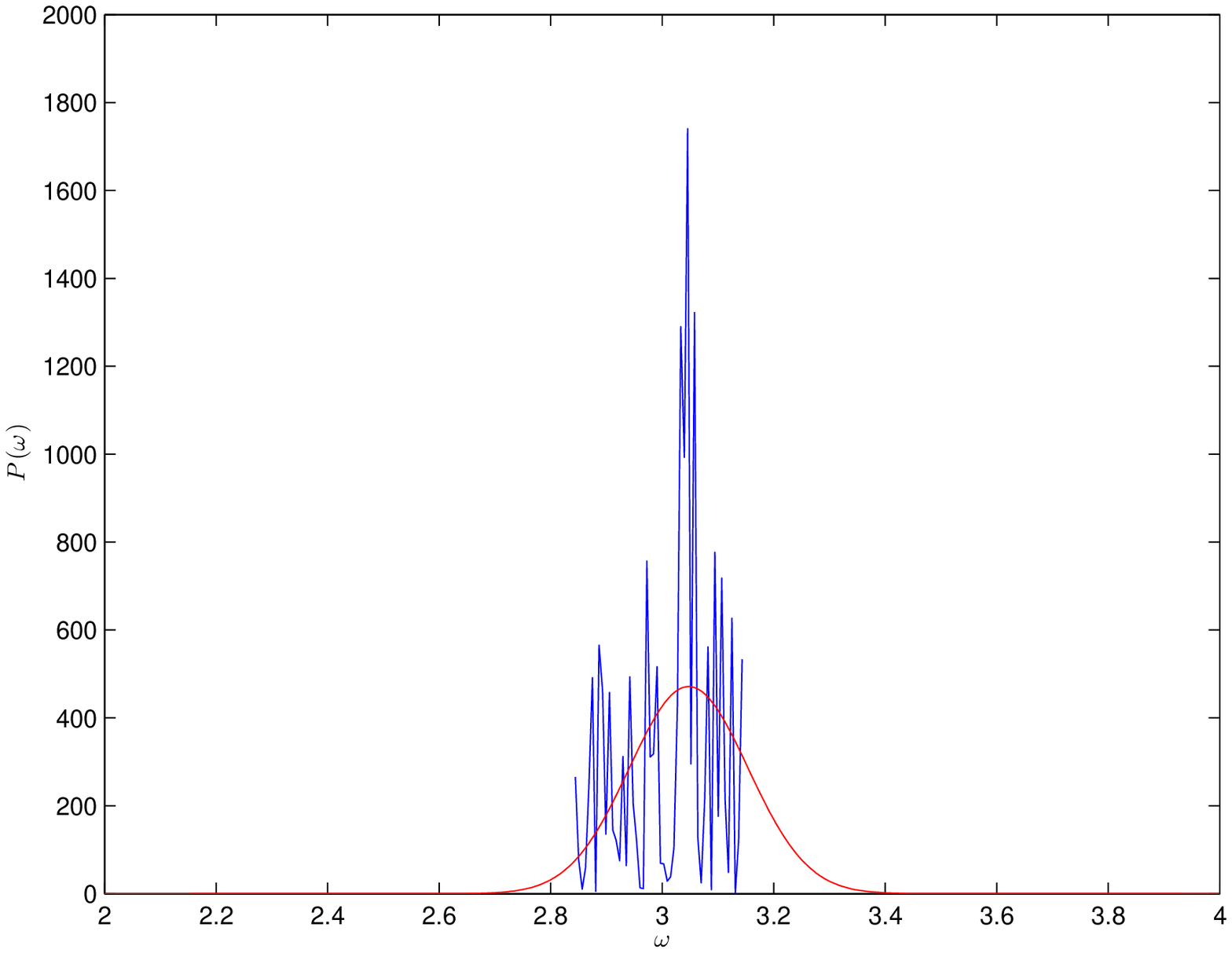}}
\quad
{\includegraphics[width=2.2in]{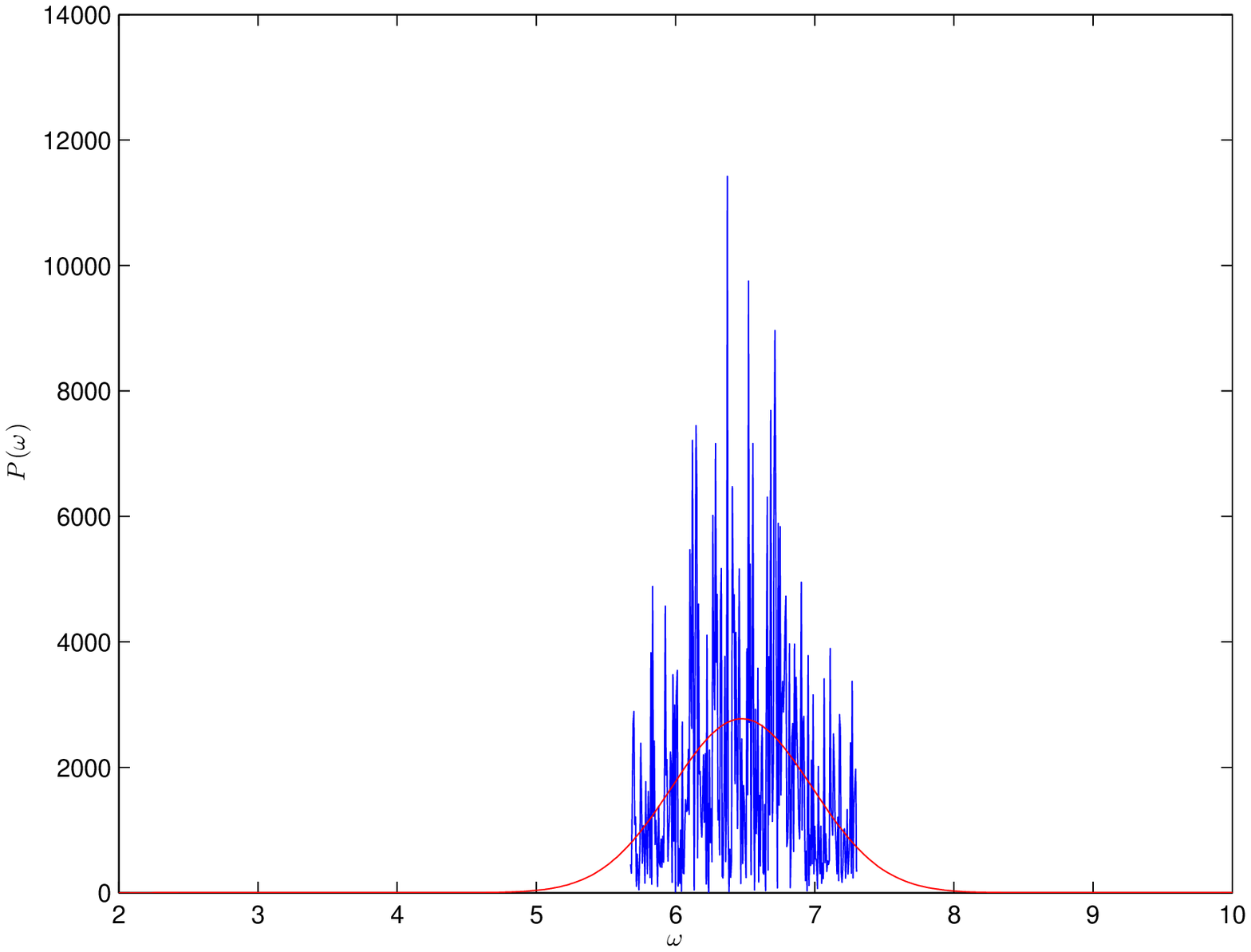}}
}
\caption{Gaussian fitted onto the 1--2pm (resting) FFT in the frequency region a) 0.07--0.4Hz, b) 2.85--3.15Hz, c) 5.68--7.3Hz.}
\label{fig:FFT-Gfit}
\end{figure}

With the availability of $24$ hour telemetry data~\cite{Nandi:12} we have monitored the variation of the dominant frequencies as a function of time. Such an analysis indicates that each hourly collected data falls into one of two categories where the FFT has a) three peaks with frequencies corresponding to $\approx 0.2$Hz, $\approx 0.3$Hz and $\approx 6$Hz, or b) two peaks with frequency values $\approx 0.03$Hz and $10$Hz. Based on this analysis itself, active and resting behaviours can be distinguished from the blood pressure data. A similar analysis (\textit{i.e.}\ Fourier transform and subsequent Gaussian fits to extract natural frequencies of physiological subsystems) was carried out on the active data set. The quality of these fits are not as good as those obtained for the resting data, and a single low frequency peak at $\approx 0.1$Hz for the resting data appears to be smeared over the range $\approx 0.1-1$Hz for the active data set. Two peak frequencies obtained from the active data set 
occur at $\omega_{4} \approx 3.3$Hz and $\approx 8.1$Hz. These values are somewhat shifted from those obtained from the resting data and raises the possibility of formulating coupled oscillator models with natural frequencies drawn from resting data, invoking a simple phenomenological form of coupling between them and attempting
to predict the resulting frequencies of the active system as a function of these coupling constants. This will be explored later.

It is to be noted however that the scope of the Fourier analysis is limited by errors occurring in the low frequency regime, arising from the finite range of collected data and phenomena that occur over long time scales. A critical analysis of such low frequencies as well as studying their diurnal variation as the system transits over various physiological states calls for a nonlinear time series analysis of the data. This is detailed in the next section.

\subsubsection{Nonlinear Analysis of Blood Pressure Data}
\label{Dimensional-Reduction}

Chaotic dynamical systems typically have attractors that have a particular structure, but within this structure, a trajectory has variation each time it goes round a cycle. Typical examples are the Lorenz attractor~\cite{Lorenz:63,Sparrow:82} and the R\"ossler attractor~\cite{Rossler:76}. So it is reasonable to ask whether the observed heart rate variability is a consequence of chaotic dynamics. There has also been much debate on this topic, with the journal {\em Chaos} addressing the issue in a series of articles under the heading {\em Introduction to Controversial Topics in Nonlinear Science: Is the Normal Heart Rate Chaotic?} \cite{Glass:09}.

The strict definition of chaos (Hasselblatt and Katok~\cite{Hasselblatt:03}) is that the system:
\begin{enumerate}
\item
must be sensitive to initial conditions (paths diverge from arbitrarily close starting points)
\item
must be topologically mixing, and
\item
its periodic orbits must be dense.
\end{enumerate}

The heart rate, or period, appears to obey some of these properties, but there is limited evidence that it is sensitive to initial conditions \cite{Kanters:94,Baillie:09}. Many of the articles in the \textit{Chaos} special issue suggest that the influence of other systems such as respiration \cite{Buchner:09,Wessel:09}, or nervous system stimulation \cite{Zhang:09} may explain most, if not all, of the heart rate variability more appropriately than any intrinsic chaotic dynamics \cite{Freitas:09}. This is in keeping with our modelling work, where we model the heart rate itself as a simple oscillator which exhibits variability due to nonlinear dynamics resulting from coupling with external oscillators. A conclusive proof of whether the blood pressure data exhibits underlying chaotic dynamics requires further analysis and is beyond the scope of this initial report.

There have been previous attempts to explain heart rate variability in terms of both respiration and nervous stimulation. Ursino and Magosso~\cite{Ursino:03} modelled heart rate variability in a spatial (blood-vessel based) compartmental  ordinary differential equation model for blood pressure and flow that incorporated heart rate control mechanisms. The model includes the action of respiration on flow in certain vessels, and nervous system control of heart rate at the sinus node in response to blood pressure changes. This model was based on an earlier model of heart rate control by Ursino~\cite{Ursino:98}, as well as other mathematical models of oscillations arising from blood pressure control.

We assume that the mouse blood pressure time series data is representative of an underlying chaotic system. A useful way of understanding chaotic systems is to view their attractors in the phase space. There are many factors that influence the blood pressure, including the sympathetic nervous system, the parasympathetic nervous system, etc. However, the only time series that we have is of the single blood pressure variable and with only this single time series, it is not possible to plot the attractor in a multi-dimensional phase space.

This is precisely the situation considered by Takens~\cite{Takens:81} who showed that delay coordinates could be used to reconstruct an attractor in phase space. Takens used a time series $x(t)$ to construct an attractor in an $n$ dimensional phase space by using delayed coordinates given by $[x(t),x(t-\tau),\ldots,x(t-(n-1)\tau)]$ where $\tau>0$ is a constant time delay. For a given system, the dimension $n$ that should be used is not necessarily clear, although work has been done to estimate a minimum embedding dimension using the method of false nearest neighbours~\cite{Kennel:92,Kennel:02}. The other uncertainty in this method is in choosing the best time delay $\tau$ to use. A method to find the best choice of $\tau$ has been proposed by minimising a mutual information function \cite{Fraser:86}.

This method has been used previously to reconstruct the attractor from arterial blood pressure data and two different plots in phase space for normal and abnormal signals have been generated~\cite{NarayanDutt:99}. However, there seems to have been no further development of these ideas.

\begin{figure}
\includegraphics[width=7cm]{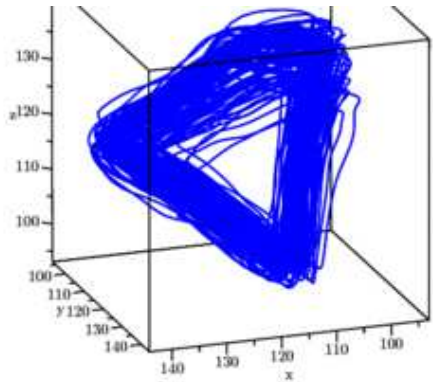}
\caption{The data in three-dimensional phase space constructed using time
delay coordinates.}
\label{phasespace}
\end{figure}

\begin{figure}[t]
\centering
\includegraphics[width=6cm]{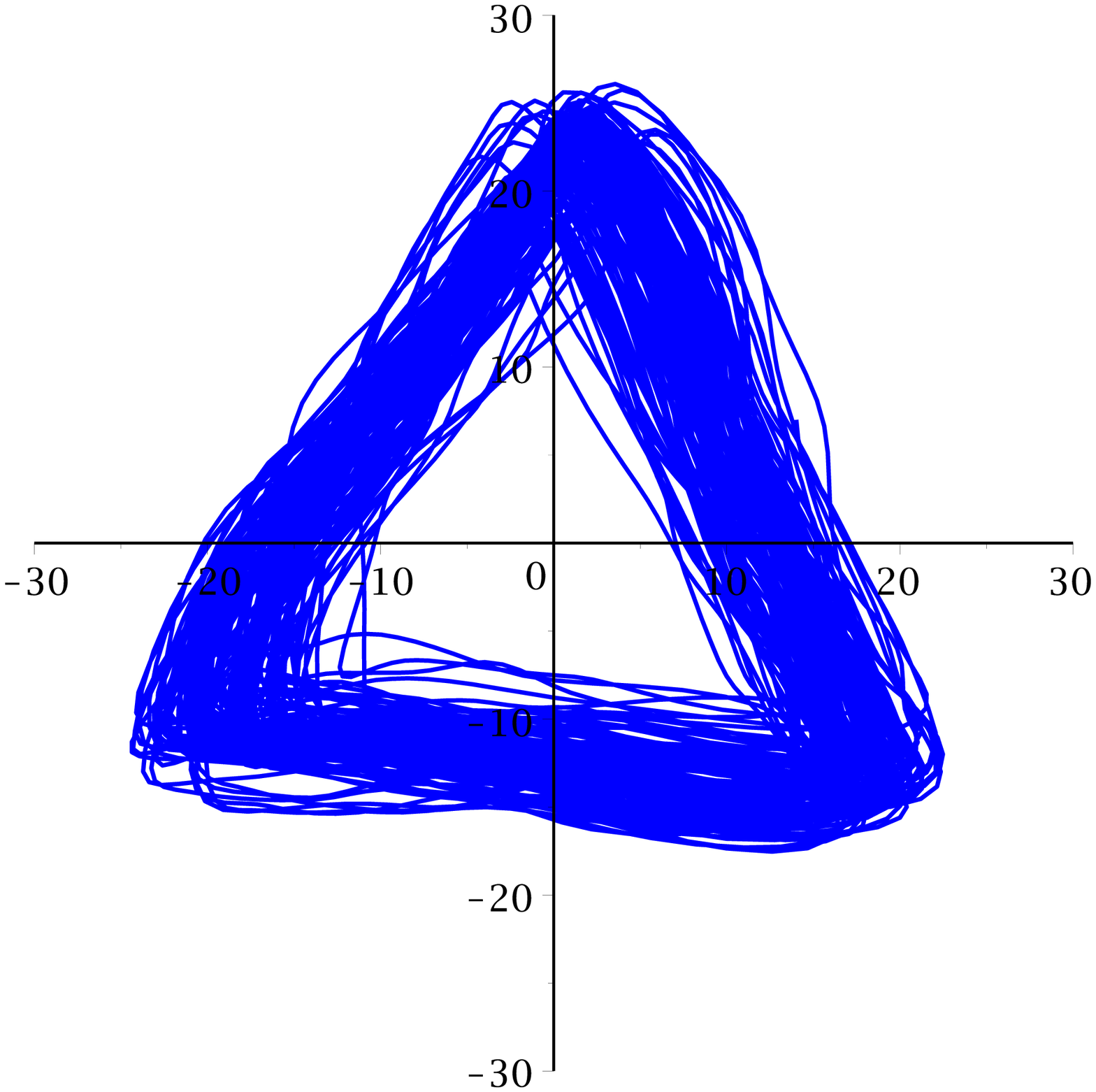}\hspace{5mm}
\includegraphics[width=6cm]{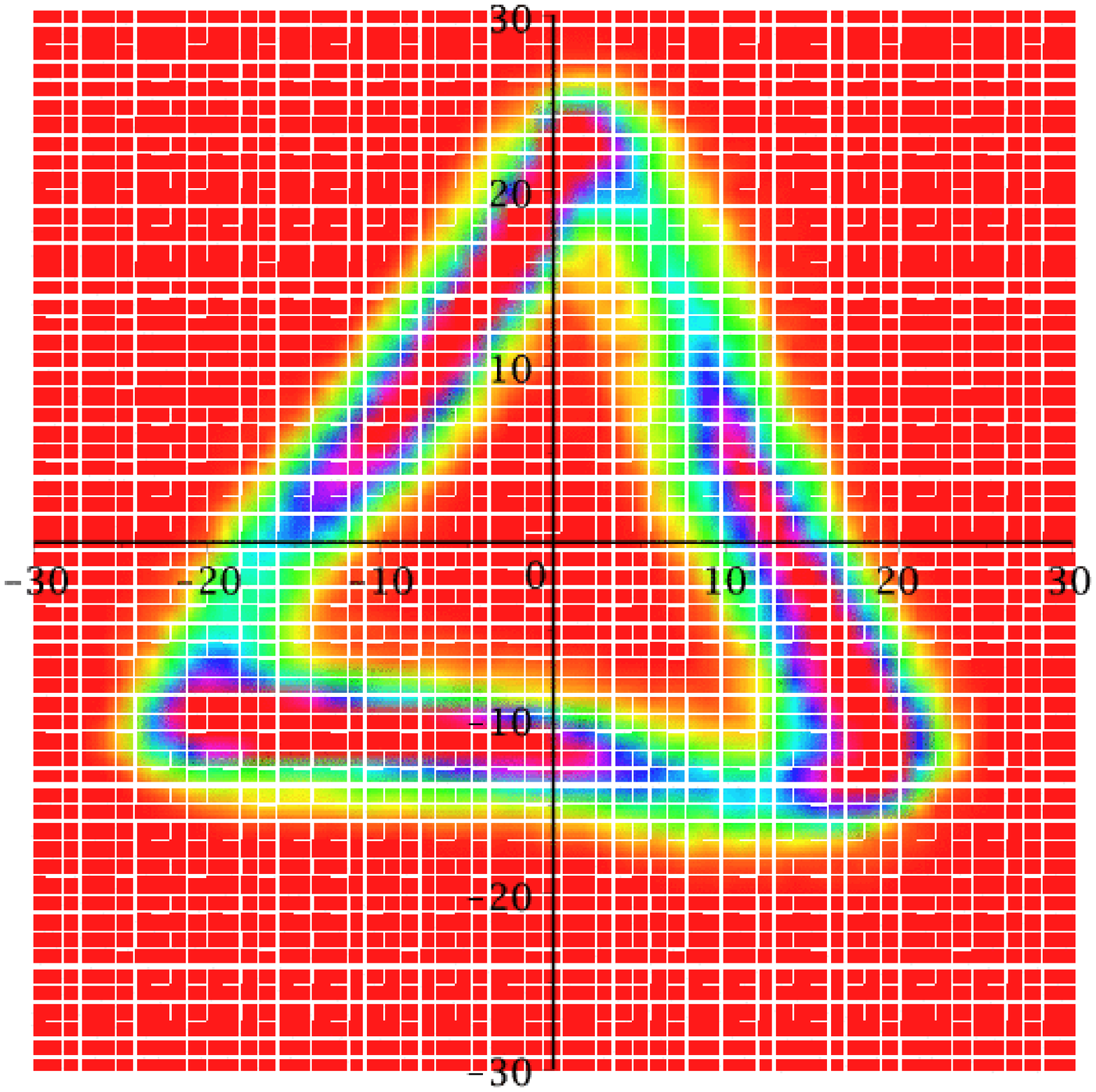}
\caption{The reconstructed trajectory projected onto a plane (left) and the density derived from the trajectory (right).}
\label{proj}
\end{figure}

We use this approach to generate an attractor in phase space for the mouse blood pressure data. For simplicity of visualisation, we choose $n=3$ and we experiment with different values of the time delay $\tau$. If $X(t)$ is the time series data, then we define the three variables $$x(t)=X(t),~~~y(t)=X(t-\tau),~~~z(t)=X(t-2\tau)$$ A plot in the three-dimensional phase space is shown in Fig.~\ref{phasespace}. By projecting this three-dimensional attractor onto a two-dimensional plane, we can catch the cycling motion of the attractor (see Fig.~\ref{proj}).

With this projection of the trajectory onto the plane, as shown in Fig.~\ref{proj}, there is simply a thick band that occurs due to the variation in each cycle. From this, it is not clear how often each part of the attractor is visited. This can be seen more clearly with a density plot in which the plane is divided up into small boxes and the number of points of the projected trajectory in each box is counted. The colour of each box is then related to the number of hits to give a density plot. The density of the trajectory in Fig.~\ref{proj} (left) is shown in Fig.~\ref{proj} (right). By computing the density over a moving window, fundamental changes in the signal can be detected.

\subsubsection{Hilbert Transform}
\label{hilbert}
The Hilbert transform can be used to extract information from an oscillating signal. The transform itself is a linear functional, mapping functions to functions. Mathematically, the Hilbert transform of a function $f(t):\mathbb{R}\mapsto\mathbb{R}$ is given by
\begin{equation}
g(t)=\frac1{\pi}\int_{-\infty}^{\infty}\frac{f(\tau)}{t-\tau}~d\tau,
\end{equation}
where the Cauchy principal value of this integral is taken. In addition, $g(t)$ is the imaginary part of the function $F(t):\mathbb{R}\mapsto\mathbb{C}$, which has real part $f(t)$ on the real axis, and which is also complex analytic in the upper half of the complex plane (if this function exists), that is $F(t)=f(t)+ig(t)$. For example, the Hilbert transform of $f(t)=A\cos(\omega(t-t_0))$ is $g(t)=A\sin(\omega(t-t_0))$, and the complex function is $F(t)=f(t)+ig(t)=Ae^{i\omega(t-t_0)}$.

By taking the amplitude and rate of change of the phase of the complex function, the Hilbert transform can be straightforwardly used to recover the amplitude and frequency of the original signal as a function of time. In the case of the example $f(t)=A\cos(\omega(t-t_0))$, the amplitude and frequency turn out to be $A$ and $\omega$, respectively, as expected.

In the case of a signal $f(t)$ that is oscillating, but not perfectly sinusoidal, the amplitude and rate of change of phase of the complex function $F(t)$ give us time varying signals that we could identify as being the amplitude and frequency of the oscillations, and may be easier to analyse than the original signal.

In this project, we endeavour to use the data to detect physiological changes, such as the difference between waking and sleeping, exercise and rest, and, ultimately, detect the possible onset of septic shock from our analysis of the data. In this section, our aim is to do this by extracting variables from the Hilbert transform that clearly indicate and quantify these changes.

\begin{figure}
\begin{center}
(a)\includegraphics[width=10cm]{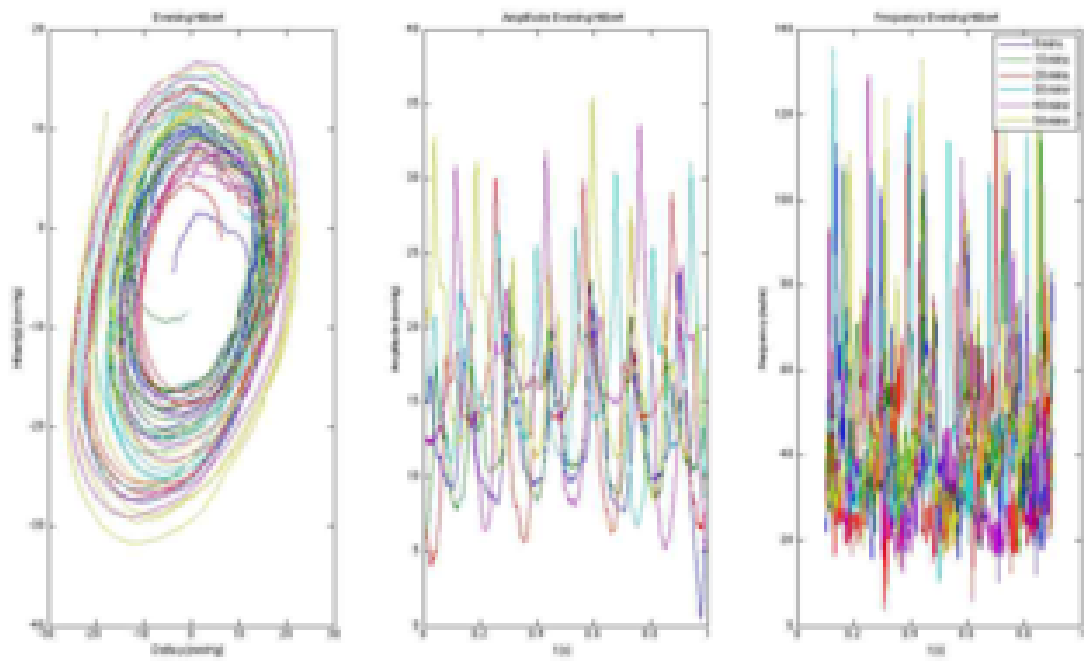}\\
(b)\includegraphics[width=10cm]{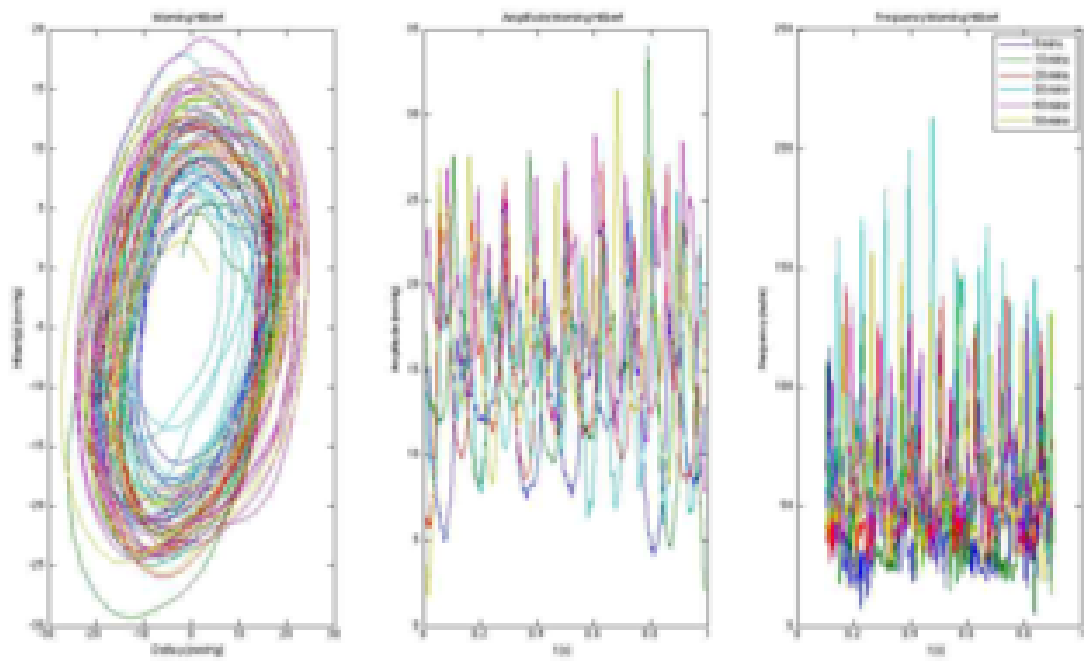}
\end{center}
\caption{Hilbert transform of measured blood pressure data. Each coloured graph represents 1s worth of data, and the different graphs are based on data spaced by intervals of 10~mins. Part (a) is data from the evening, and (b) is from the morning. In both cases, the left-hand sub-figure shows the Hilbert transform against the original data (minus mean), the central figure shows the amplitude variation in time, and the right-hand figure shows the frequency variation, calculated as the time-derivative of the phase.}
\label{fig:hilbert}
\end{figure}

As with the Fourier transform, the Hilbert transform can also be used to extract amplitude and frequency data from a signal. However, the Hilbert transform has the advantage that nonlinear data can be analysed. Unlike the Fourier transform, which analyses the signal by breaking it down into its component harmonics, the Hilbert transform expresses the instantaneous frequency as a function of time. So, for example, a single ectopic beat in an otherwise fairly periodic signal would cause a large change in the resulting Fourier transform, but with the Hilbert transform, the disturbance remains localised in time (likewise if a small period of data is omitted, due to experimental error, for example).

We used the pressure data, collected at $1000$Hz, subtracted the localised mean, and then computed the Hilbert transform. This was plotted against the original signal (minus the mean), and the results are shown in Fig.~\ref{fig:hilbert}. As can be seen, the result is a time-varying indication of amplitude and frequency of the signal. However, the resulting waveforms are still complicated, and might not indicate the desired physiological changes sufficiently clearly.

One possible way to reduce the complexity is to perform the empirical mode decomposition repeatedly, until an intrinsic mode function is obtained. In this way, we would expect to be able to analyse data over a longer period, such as an hour. See~\cite{huang-shen:05} for more detail on the use of the empirical mode decomposition. As a result the physiological changes would be expected to be more evident. This will be the subject of future work.

\subsection{Linear, Nonlinear and Forced Coupled Oscillator Models}
\label{diffmodels}

The starting point of modelling cardiovascular variability is embroiled in the interdependencies of the various physiological subsystems indicated in Fig.~\ref{Figure1}. Starting with the simplest assumption of a set of coupled linear oscillators the collective dynamics of the system can be written as
\begin{eqnarray}\label{eq:coup-lin-osc}
\ddot{x}_{1} + \omega^{2}_{1} x_{1} + k_{12} \left(x_{1} - x_{2} \right) + k_{13} \left(x_{1} - x_{3} \right) &+& \nonumber \\ k_{14} \left(x_{1} - x_{4} \right) &=& f_{0} \cos(\omega_f t), \nonumber \\
\ddot{x}_{2} + \omega^{2}_{2} x_{2} + k_{12} \left(x_{2} - x_{1} \right) + k_{24} \left(x_{2} - x_{4} \right) &=& 0, \nonumber \\
\ddot{x}_{3} + \omega^{2}_{3} x_{3} + k_{13} \left(x_{3} - x_{1} \right) &=& 0, \nonumber \\
\ddot{x}_{4} + \omega^{2}_{4} x_{4} + k_{24} \left(x_{4} - x_{2} \right) + k_{14} \left(x_{4} - x_{1} \right) &=& 0,
\end{eqnarray}
where $x_{1}$, $x_{2}$, $x_{3}$ and $x_{4}$ correspond to the heart rate (HR), sympathetic nervous system (SNS), parasympathetic nervous system (PNS) and respiratory system (RS) respectively. The natural frequencies of each of the oscillators are described by $\omega_{i}$, while $k_{ij}$ denotes the coupling between the different physiological subsystems. In our attempt to model real physiological systems a forced coupled linear oscillator corresponding to the heart was used with the forcing having an amplitude $f_0$ with a frequency $\omega_{f}$.

We solved the coupled set of unforced ($f_{0}=0$) and forced ($f_{0} \neq 0$) oscillator equations numerically using the Matlab ODE suite as a function of the parameters $\omega_{i}$ ($i = 1, 2, \ldots 4$), and $k_{12}$, $k_{13}$, $k_{14}$, and $k_{24}$. The coupling constants between SNS and PNS $k_{23}$, and between PNS and respiratory system RS $k_{34}$, are chosen to be zero as guided by an initial review of the physiological literature.

The natural frequencies of the oscillators were obtained by Fourier transforming the resting blood pressure data and fitting a Gaussian function to extract the mean and variance near each local maxima. The mean corresponds to the natural frequencies of the oscillators, $\omega_{1} \approx 7$Hz, $\omega_{2} \approx 0.1$Hz, $\omega_{3} \approx 1$Hz, and $\omega_{4} \approx 3$Hz respectively. The obtained frequencies agree with those obtained via other physiological methods. We have varied these parameters over the range $\omega_{i} \pm \Delta \omega_{i}$, where $\Delta \omega_{i}$ is the variance of each of the Gaussian functions fitted to the Fourier transformed data to look for qualitative differences in the time series data for each of the oscillators.

\begin{figure}
\includegraphics[width=10cm]{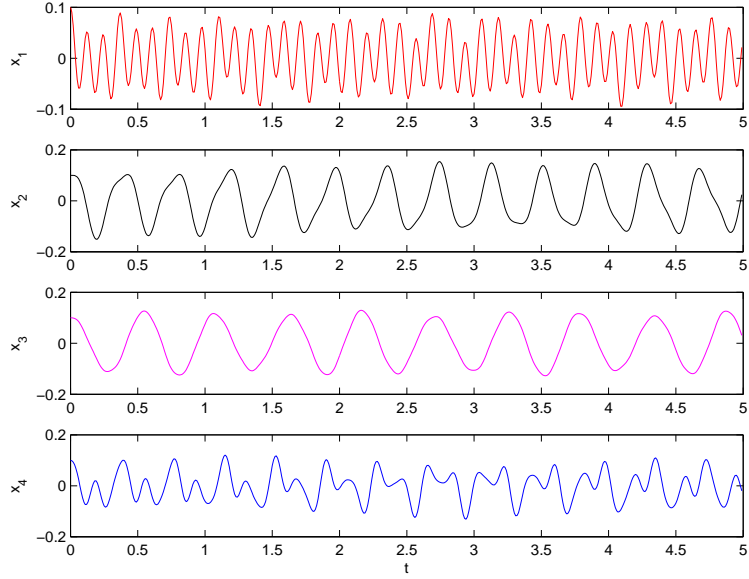}
\caption{Numerical solution of the physiological system modelled as a set of coupled linear oscillators (see Eq.~\ref{eq:coup-lin-osc}). The panels show time variation of the various physiological subsystems HR ($x_{1}$), SNS ($x_{2}$), PNS ($x_{3}$) and RS ($x_{4}$). The natural frequencies of the oscillators are assumed to be $\omega_{1} = 7$Hz, $\omega_{2} = 0.1$Hz, $\omega_{3} = 1$Hz, $\omega_{4} = 3$Hz and coupling constants $k_{12} = 100$Hz$^2$, $k_{13} = 100$Hz$^2$, $k_{14} = 400$Hz$^2$, $k_{24} = 300$Hz$^2$.}
\label{couplinosc.eps}
\end{figure}

\begin{figure}
\includegraphics[width=8cm]{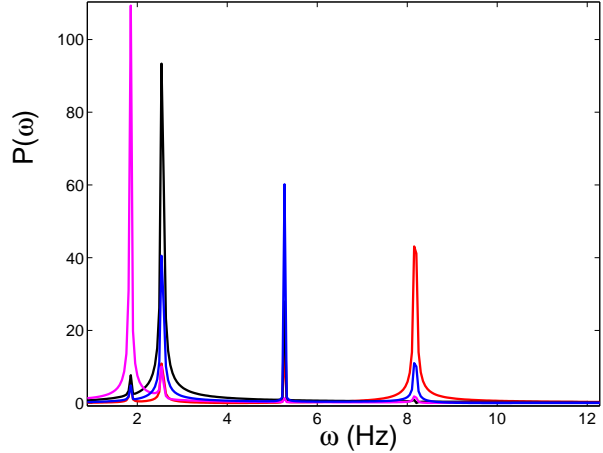}
\caption{Fourier transform of the time series data obtained by numerical solution of the coupled oscillator system (see Eq.~\ref{eq:coup-lin-osc}). The frequencies of the resulting system that are different from the natural oscillator frequencies can be seen.}
\label{couplinoscfreq}
\end{figure}

In the absence of experimental values for the coupling constants $k_{ij}$, we have varied them over several orders of magnitude to look for qualitative and quantitative effects on the generated time series of the various oscillators. A representative time series corresponding to $\omega_{i}$ values quoted above (\textit{i.e.}\ $\omega_{1} = 7$Hz, $\omega_{2} = 0.1$Hz, $\omega_{3} = 1$Hz, and $\omega_{4} = 3$Hz) and coupling constants $k_{12} = 100$Hz$^2$, $k_{13} = 100$Hz$^2$, $k_{14} = 400$Hz$^2$ and $k_{24} = 300$Hz$^2$ is shown in Fig.~\ref{couplinosc.eps}.
The wave train can be Fourier transformed to generate a frequency spectrum corresponding to the combined oscillator system. This can be seen in Fig.~\ref{couplinoscfreq}

It should be be noted that the data train generated using the linear oscillator model does not look as ``chaotic'' as the experimental blood pressure data (see Fig.~\ref{TimeSeries}) and blown up section (see Fig.~\ref{data}). However, the main motivation of invoking a coupled linear oscillator model is not to match the exact blood pressure waveform data but rather to put bounds on the coupling constants that can be tested via physiological measurements. In fact the true power of this approach lies in the ability to derive a closed form expression of the resulting frequencies of the coupled linear oscillator system as a function of the model constant $\omega_{i}$ and $k_{ij}$. Demanding that the square of the frequency is real and positive automatically imposes bounds on the value of the coupling constants.

In order to generate oscillatory behaviour of a coupled oscillator system that resembles the physiological data more closely we use a variant of the model proposed by Stefanovska \textit{et al.}~\cite{Stefanovska:01,Stefanovska:01a}. In this model the $i^{\rm th}$ oscillator is represented by the set of dynamical equations
\begin{eqnarray}\label{eq:coup-nonlin-osc}
\dot{x}_{i} &=& - x_{i} q_{i} - \omega_{i} y_{i}, \nonumber \\
\dot{y}_{i} &=& - y_{i} q_{i} + \omega_{i} x_{i},
\end{eqnarray}
where $q_{i} = \alpha_{i} \left( \sqrt{x^{2}_{i} + y^{2}_{i}} - a_{i} \right)$, $a_{i}$ denotes the amplitude of each oscillator, $\omega_{i}$ corresponds to the natural frequencies of the oscillator, while the constant $\alpha_{i}$ determines the stability of the limit cycle the model admits.

\begin{figure}
\includegraphics[width=10cm]{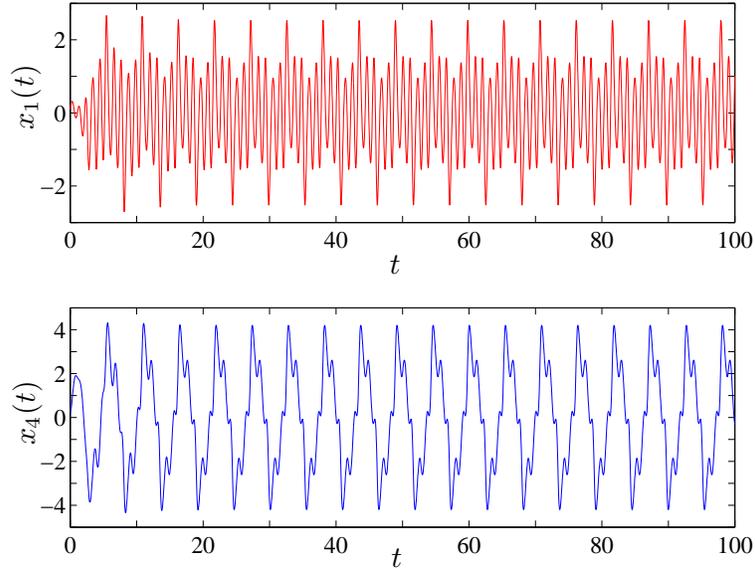}
\caption{Time series data of the two coupled nonlinear oscillators obtained via numerical solution of Eq.~\ref{eq:coup-nonlin-osc}. The parameter values for the data shown are $\omega_{1}=7$Hz, and $\omega_{4}=2$Hz, the oscillator amplitudes $a_1 = 1$, and $a_2 = 0.7$ and the parameters determining the stability of the limit cycles are $\alpha_{1} = \alpha_{3} = 1$.}\label{twocoup-nonlinosc}
\end{figure}
\begin{figure}
\includegraphics[width=8cm]{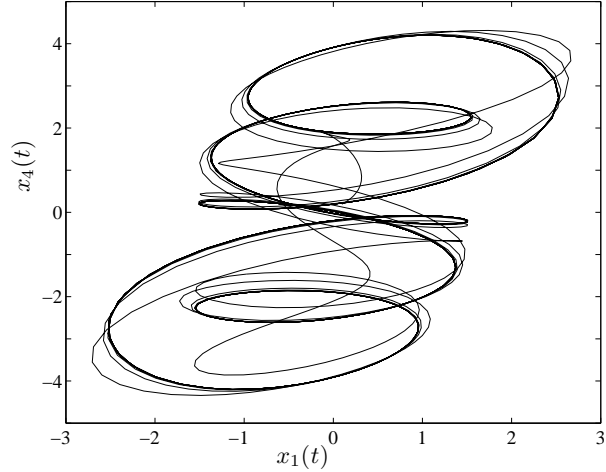}
\caption{Phase portrait of the time series data for the HR and RS system as shown in Fig.~\ref{twocoup-nonlinosc}.}\label{twocoup-nonlinosc-pspace}
\end{figure}

Similar to our linear oscillator model we propose a minimal coupling between oscillators $i$ and $j$ of the form $k_{ij} \left(x_{i} - x_{j} \right)$. A semi-microscopic justification of such a coupling between the respiratory and the cardiovascular oscillators has been developed by Stefanovska \textit{et al.}~\cite{Stefanovska:01,Stefanovska:01a}. For a set of parameter values this coupled nonlinear oscillator model can be solved numerically using the Matlab ODE suite. The resulting data set for a coupled cardiovascular (HR) and respiratory system (RS), corresponding to $x_{1}$ and $x_{4}$ is shown in Fig.~\ref{twocoup-nonlinosc}. The nonlinear oscillator model and its numerical solution further allows for the exploration of the nature of the dynamical system by looking at attractors in phase space as shown in Fig.~\ref{twocoup-nonlinosc-pspace}.

This provides a route to a three way synergistic investigation of the original blood pressure data by a combination of physiological experiments, data analysis \textit{viz.}\ Fourier and Hilbert transforms and attractor reconstructions, and numerical solutions of linear and nonlinear coupled oscillator models. We believe that a minimal model of coupled oscillator models that quantitatively describes the complex physiological data could be developed via such a method. Further investigations of each of these techniques will be explored in separate publications.

\begin{figure}
\includegraphics[width=10cm]{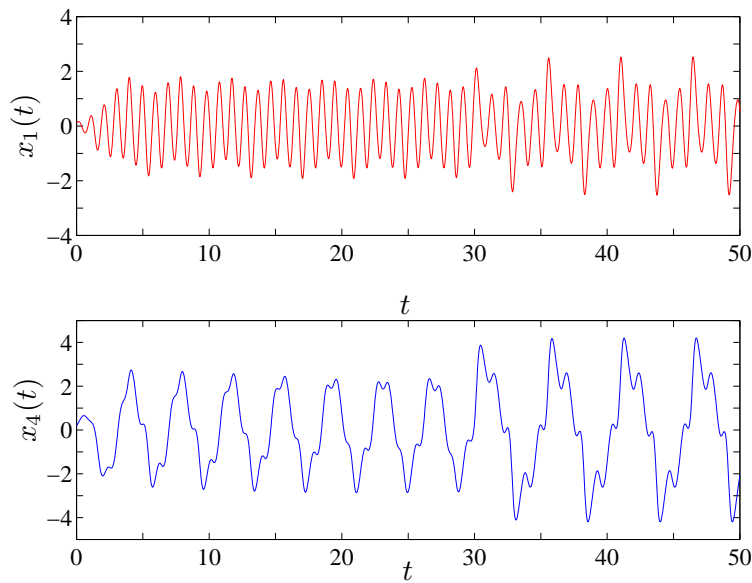}
\caption{The HR and RS system modelled as two coupled oscillators similar to Stefanovska~\cite{Stefanovska:01,Stefanovska:01a}, with a linear coupling between oscillators (see text). The coupling constant between the oscillators undergoes a step change from a value of $k_{14}=2$ to $k_{14}=3$ at time $t=30$. A reduction in frequency is seen in both the HR and RS systems $x_{1}$ and $x_{4}$. The natural frequencies of the oscillators for the data shown are $\omega_{1} \approx 7$Hz, and $\omega_{4} \approx 2$, the oscillator amplitudes are $a_1 = 1$ and $a_2 = 0.7$ respectively. The parameters determining the stability of the limit cycles are $\alpha_{1} = \alpha_{3} = 1$.}
\label{shock-twocoup-nonlinosc}
\end{figure}

The linear and nonlinear coupled oscillator model allows us to further quantitatively address the original hypothesis, \textit{i.e.}\ a reduction of the coupling between the different oscillators leads to decreased HRV in septic shock. Physiological signatures suggest that the coupling between the HR and RS systems is the most dominant. We have thus modelled a coupled HR and RS system where the coupling constant $k_{14}$ undergoes a step change from a base value of $k_{14} = 2$Hz$^2$ to $k_{14} = 3$Hz$^2$ at a time $t = 30$. This is shown in Fig.~\ref{shock-twocoup-nonlinosc}.

\section{Discussion and Future directions}
\label{future}
In summary, we have developed a framework to study HRV in conscious mouse blood pressure data via a synergistic experiment, theory and data analysis approach. Blood pressure data collected over a $24$ hour period have been analysed using Fourier and Hilbert transforms and nonlinear data analysis techniques such as attractor reconstruction. Our initial starting point was that the complex blood pressure data is a result of the coupling between four physiological subsystems that are modelled as oscillators: the respiratory system (RS), the cardiovascular system (HR) and the sympathetic (SNS) and parasympathetic nervous systems (PNS). The natural frequencies of these oscillators have been extracted from the Fourier transform data and are in good agreement with values quoted in the literature~\cite{Baudrie:07}. Further the Hilbert transform and the attractor reconstruction techniques applied to the physiological data provides us with the ability to quantitatively distinguish between active and resting behaviour
of mice. Lastly the theoretical analysis of linear and nonlinear coupled oscillator models helps us identify different dynamical scenarios as a function of the parameters of the model, namely the natural frequencies of the four physiological oscillator subsystems and their coupling constants. It also allows us to quantify changes in heart rate variability (HRV) as a function of a step change in any of the coupling constants. It is believed that such a feature might be operative when the mouse is subject to septic shock.

The power of this approach is that linear modelling of oscillator systems allows the design of experiments to both test the mathematical model and determine the intrinsic oscillator frequencies and coupling constants between oscillators. Forcing one oscillator into a different intrinsic frequency would cause changes in the measured frequencies of the other oscillators and would allow calculation of the coupling constants between oscillators. Similarly, pharmacological modulation of a coupling constant would allow estimation of intrinsic oscillator frequencies. A combination of both approaches would be required in order to fully describe the intrinsic frequencies and coupling constants between the oscillators.


\section{3Rs Impact}

The data currently utilised has been obtained by radiotelemetry -- a gold standard technique for monitoring cardiovascular parameters from conscious freely roaming animals within their home cages. This does not necessitate restraint or investigator intervention and this in itself reduces stress and suffering experienced by animals.
Large, physiologically relevant data sets can be obtained from relatively small numbers of animals compared with traditional techniques (such as tethering; which introduces a significant stress component, high drop out rate and increased inter animal variability within data sets). Given the low inter-animal data variability, power calculations of our data have revealed that around 8 animals are required (per group) to detect differences in cardiovascular parameters, when comparing wild type verses GM or pharmacologically treated animals in non-diseased conditions. This is in contrast to around 20 animals when using tethered systems. We and others have also achieved a $90$-$100$\% recovery rate in all animals implanted with radiotelemetry probes.

As this project aims to improve methods for analysing and interpreting large data sets, it would enable re-analysis of existing electronic data sets without requiring any further animal experimentation. We have already obtained an expression of interest for the re-analysis of historical data sets held within the pharmaceutical industry and other academic laboratories. This analysis may reveal otherwise unidentified biological phenomena that could contribute towards the understanding of how certain gene products or pharmacological interventions impact upon cardiovascular function. The analysis of data obtained from multiple species would also determine the robustness of the developed mathematical models and applicability to human data.

This method may also allow refinement of current animal models of disease (such as sepsis which causes moderate to severe suffering) by aligning the endpoints towards those used in the clinic (\textit{e.g.}\ changes in heart rate variability, oscillator uncoupling) which are likely to occur at earlier stages. This would provide a predictive tool for assessing the risk of onset of septic shock, and away from more commonly used severe endpoints such as organ failure and survival.

Finally, this project will collate historical data sets providing a repository that can be accessed by other investigators and may avoid the need to undertake further animal experimentation.

\section{Acknowledgements}
The authors acknowledge support from the EPSRC and NC3Rs via the Mathematics in Medicine Study Group.

\end{document}